# Employment of Jacobian elliptic functions for solving problems in nonlinear dynamics of microtubules

Slobodan Zeković[a], Annamalai Muniyappan[b], Slobodan Zdravković[a,*], Louis Kavitha[b,c]

[a] *Institut za nuklearne nauke Vinča, Laboratorija za atomsku fiziku 040, Univerzitet u Beogradu, Poštanski fah 522, 11001 Beograd, Serbia*
[b] *Department of Physics, Periyar University, Salem-636 011, Tamilnadu, India*
[c] *Center for Nanoscience and Nanotechnology, Periyar University, Salem-636 011, Tamilnadu, India*

*Keywords:*
Jacobian elliptic functions
Ordinary differential equations
Microtubules
Kink soliton

A B S T R A C T

We show how Jacobian elliptic functions (JEF) can be used to solve ordinary differential equations (ODE) describing nonlinear dynamics of microtubules (MT). We demonstrate that only one of JEFs can be used while the remaining two do not represent the solutions of the crucial differential equation. We show that a kink-type soliton moves along MT. Beside this solution, we discuss a few more that may or may not have physical meaning. Finally, we show what kinds of ODE can be solved using JEFs.

## 1. Introduction

Microtubules (MTs) are important cell proteins. They represent cytoskeletal structures and serve as a "road network" for motor proteins (kinesin and dynein) dragging different molecular cargos. They are hollow tubes with inner and outer diameter of 17nm and 25nm [1]. Its length may span dimensions from the order of micrometer to the order of millimetre. The cylindrical wall is usually formed by 13 longitudinal structures called protofilaments (PFs), as shown in Fig. 1. They are series of proteins known as tubulin dimers [2,3]. Each dimer is an electric dipole whose length and longitudinal component of the electric dipole moment are $l = 8$nm [2-4] and $p = 337$Debye [5,6], respectively.

---

* Corresponding author.
  *E-mail address:* szdjidji@vinca.rs.

The constituent parts of the dimers are $\alpha$ and $\beta$ tubulins, corresponding to positively and negatively charged sides, respectively [2-4].

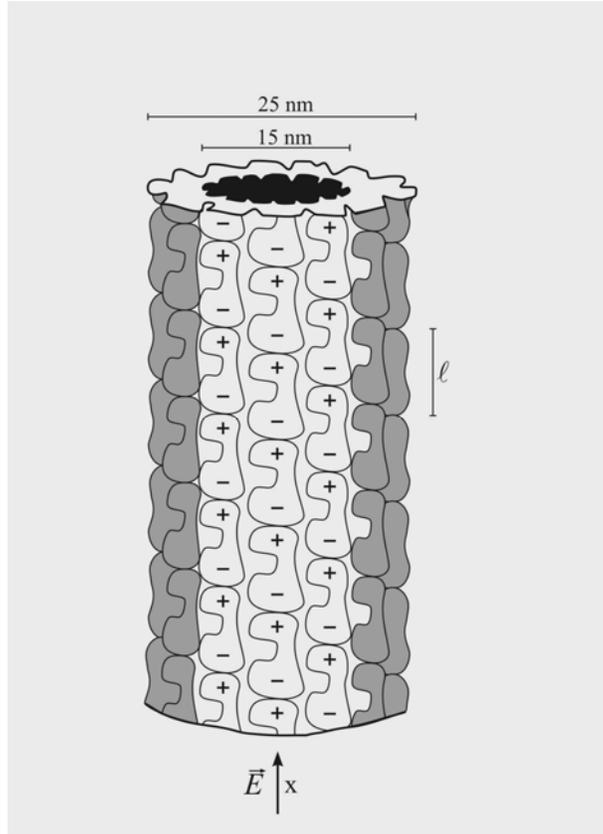

**Fig. 1.** The structure of a microtubule

This paper is organized as follows. In Section 2, we briefly review one of the models describing MT dynamics. It is well known that, for a travelling wave, a partial differential equation (PDE) can be transformed into ODE. This equation, crucial for the model, is solved in Section 3 using JEFs. We do believe that this method is the simplest and the most elegant of all. A couple of solutions of the mentioned equation that may or may not have physical meaning are studied in Section 4. In Section 5 we show what kind of ODEs can be solved using JEFs. Finally, concluding remarks are in Section 6.

**2. Z-model of MTs**

The model assumes only one degree of freedom per dimer. This is $z_n$, a longitudinal displacement of a dimer at a position n in a direction of PF, which is a direction of x-axes. Hence, we refer to the model as z-model [7]. In what follows we very briefly describe the model. Of crucial importance is the fact that the bonds between dimers within the same PF are much stronger than the soft bonds between neighbouring PFs [8,9]. This implies that the longitudinal displacements of pertaining dimers in a single PF should cause the longitudinal wave propagating along PF. All this means that we can write Hamiltonian for one PF only and this is [7]

$$H = \sum_n \left[ \frac{m}{2} \dot{z}_n^2 + \frac{k}{2}(z_{n+1} - z_n)^2 + V(z_n) \right], \quad (1)$$

where dot means the first derivative with respect to time, $m$ is a mass of the dimer and $k$ is a harmonic constant describing the nearest–neighbour interaction between the dimers belonging to the same PF. The first two terms are kinetic energy of the dimer and harmonic energy, respectively. The latter one is an elastic energy, which is a potential energy of the chemical interaction between the neighbouring dimers belonging to the same PF. The last term in Eq. (1) is the combined potential

$$V(z_n) = -Cz_n - \frac{1}{2} A z_n^2 + \frac{1}{4} B z_n^4, \quad C = qE, \quad (2)$$

where $E$ is the magnitude of the intrinsic electric field while $q$ is the excess charge within the dipole. It is assumed that $q > 0$ and $E > 0$. The first term in Eq. (2) is an energy of the dimer at the site $n$ in the field $E$ and the remaining two represent the well known double-well potential with positive parameters $A$ and $B$ that should be determined [10]. The double-well potential is rather common in physics [11-14]. The first attempt to use it in nonlinear dynamics of MTs was done about 20 years ago [10,15].

From Eqs. (1) and (2), assuming a continuum approximation $z_n(t) \to z(x,t)$, we straightforwardly obtain the following nonlinear PDE of motion

$$m \frac{\partial^2 z}{\partial t^2} - k l^2 \frac{\partial^2 z}{\partial x^2} - qE - Az + Bz^3 + \gamma \frac{\partial z}{\partial t} = 0, \quad (3)$$

where the last term represents a viscosity force with $\gamma$ being a viscosity coefficient [7,10]. For a travelling wave $z(x,t) = z(\xi)$ the PDE (3) transforms into ordinary differential equation [7,10]

$$\alpha \psi'' - \rho \psi' - \psi + \psi^3 - \sigma = 0, \quad (4)$$

where $\psi' \equiv d\psi/d\xi$,

$$\alpha = \frac{m\omega^2 - kl^2\kappa^2}{A}, \quad \rho = \frac{\gamma\omega}{A}, \quad \sigma = \frac{qE}{A\sqrt{\frac{A}{B}}} \quad (5)$$

and

$$z = \sqrt{\frac{A}{B}}\psi. \qquad (6)$$

The unified variable $\xi$ is

$$\xi = \kappa x - \omega t, \qquad (7)$$

where $\kappa$ and $\omega$ are constants.

A standard way for solving Eq. (4) is rather tedious [10,12]. More elegant and relatively new procedure is modified extended tanh-function (METHF) method [16-21]. In the next section we demonstrate elegance of the method based on JEFs.

### 3. Jacobian elliptic functions

As was stated above, the main purpose of this work is to study if and how JEFs can be used to solve biophysical problems. In particular, we demonstrate the usefulness of these functions in solving Eq. (4).

Properties of JEFs $\mathrm{sn}(\beta\xi)$, $\mathrm{cn}(\beta\xi)$ and $\mathrm{dn}(\beta\xi)$ with the modulus $m$ can be found in many textbooks and papers [22-26]. A basic algebra includes the following formulae:

$$\mathrm{sn}^2(x) + \mathrm{cn}^2(x) = 1, \qquad \mathrm{dn}^2(x) = 1 - m^2\,\mathrm{sn}^2(x), \qquad (8)$$

and

$$\mathrm{sn}'(x) = \mathrm{cn}(x)\,\mathrm{dn}(x), \quad \mathrm{cn}'(x) = -\mathrm{sn}(x)\,\mathrm{dn}(x), \quad \mathrm{dn}'(x) = -m^2\,\mathrm{sn}(x)\,\mathrm{cn}(x), \qquad (9)$$

where prime denotes the first derivative. When $m \to 1$, the JEFs degenerate into hyperbolic functions, that is

$$\mathrm{sn}(x) \to \tanh(x), \qquad \mathrm{dn}(x) \to \mathrm{sech}(x), \qquad \mathrm{cn}(x) \to \mathrm{sech}(x). \qquad (10)$$

Also, for $m \to 0$ these functions degenerate into trigonometric functions but this is not relevant for this paper.

Our ansatz is the following function

$$\psi = a_0 + \sum_{i=1}^{M}\left(a_i\Phi^i + b_i\Phi^{-i}\right), \qquad (11)$$

where $\Phi = \Phi(\xi)$ is one of JEFs. We start with the function

$$\Phi = \mathrm{sn}(\beta\xi), \qquad (12)$$

where $\beta$ is a parameter to be determined. The highest orders of the function $\Phi$ in the expressions for $\psi''$ and $\psi^3$ are $\Phi^{M+2}$ and $\Phi^{3M}$ respectively, which brings about $M=1$. Using Eqs. (8) and (9) we can easily show that the second derivative $\psi''$ can be expressed in terms of $\Phi$, while the first one is proportional to $\text{cn}(\beta\xi)\text{dn}(\beta\xi)$. This term can be expressed through $\Phi$ only for $m=1$. Hence, for $M=1$ and $m=1$ we come up with

$$\psi = a_0 + a_1\Phi + b_1\Phi^{-1}, \tag{13}$$

$$\psi' = \beta\left(a_1 - a_1\Phi^2 + b_1 - b_1\Phi^{-2}\right) \tag{14}$$

and

$$\psi'' = 2\beta^2\left(a_1\Phi^3 - a_1\Phi - b_1\Phi^{-1} + b_1\Phi^{-3}\right), \tag{15}$$

where.

$$\Phi = \tanh(\beta\xi), \tag{16}$$

which comes from Eqs. (10) and (12).

If we plug the expressions for $\psi$, $\psi'$, $\psi''$ and $\psi^3$ into Eq. (4) we obtain the crucial equation

$$A_1\Phi + B_1\Phi^{-1} + A_2\Phi^2 + B_2\Phi^{-2} + A_3\Phi^3 + B_3\Phi^{-3} + A_0 = 0, \tag{17}$$

where the following set of abbreviations is used:

$$A_0 = -a_0 + a_0^3 + 6a_0 a_1 b_1 - \rho\beta a_1 - \rho\beta b_1 - \sigma, \tag{18}$$

$$A_1 = -a_1 + 3a_0^2 a_1 + 2\alpha\beta^2 a_1 + 3a_1^2 b_1, \tag{19}$$

$$B_1 = -b_1 + 3a_0^2 b_1 - 2\alpha\beta^2 b_1 + 3a_1 b_1^2, \tag{20}$$

$$A_2 = 3a_0 a_1^2 + \rho\beta a_1, \tag{21}$$

$$B_2 = 3a_0 b_1^2 + \rho\beta b_1, \tag{22}$$

$$A_3 = 2\alpha\beta^2 a_1 + a_1^3 \tag{23}$$

and

$$B_3 = 2\alpha\beta^2 b_1 + b_1^3. \tag{24}$$

Of course, Eq. (17) is satisfied if all these coefficients are simultaneously equal to zero which yields a system of seven equations. The solutions expressed through hyperbolic cotangent cannot be biophysically tractable as this function diverges. This means that we are looking for the acceptable solutions for which $a_1 \neq 0$ and $b_1 = 0$. A more general case is studied in Section 4. Hence, we deal with the system of four equations only and its solution, i.e. the values of the parameters $\beta$, $a_0$, $a_1$ and $\alpha$, are given through

$$8a_0^3 - 2a_0 + \sigma = 0, \tag{25}$$

$$a_1^2 = 1 - 3a_0^2, \tag{26}$$

$$\rho\beta = -3a_0 a_1 \tag{27}$$

and

$$2\alpha\beta^2 = -a_1^2. \tag{28}$$

Notice that $\alpha < 0$ and that there should be

$$a_0^2 < \frac{1}{3} \tag{29}$$

for $a_1$ to be real.

The polynomial (25) has three real roots for [28]

$$\sigma < \sigma_0 = \frac{2}{3\sqrt{3}}. \tag{30}$$

These solutions, shown in Fig. 2, are

$$a_{01} = \frac{1}{2\sqrt{3}}\left(\cos F + \sqrt{3}\sin F\right), \tag{31}$$

$$a_{02} = \frac{1}{2\sqrt{3}}\left(\cos F - \sqrt{3}\sin F\right), \tag{32}$$

$$a_{03} = -\frac{1}{\sqrt{3}}\cos F, \tag{33}$$

where

$$F = \frac{1}{3}\arccos\left(\frac{\sigma}{\sigma_0}\right). \tag{34}$$

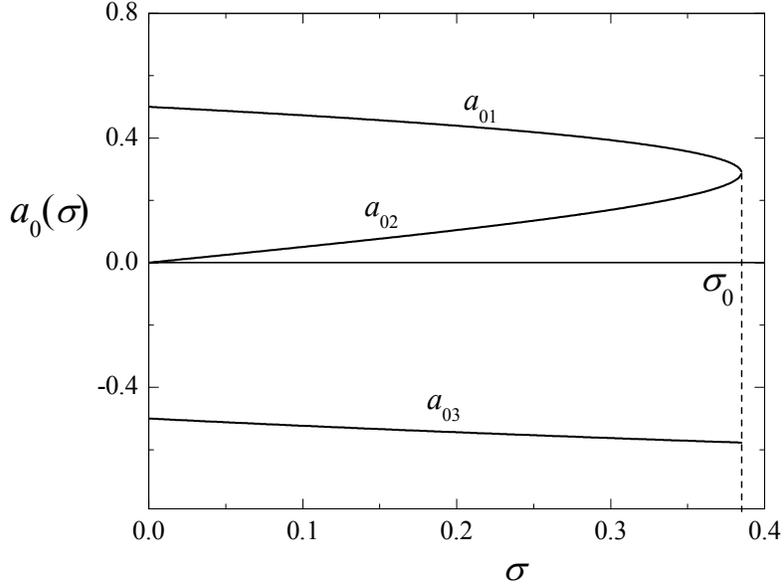

**Fig. 2.** The values of $a_0$ as a function of the parameter $\sigma$.

According to Eqs. (13), (26) and (27) we easily obtain the following solutions of Eq. (4) for $b_1 = 0$

$$\psi_k(\xi) = a_{0k} - \sqrt{1 - 3a_{0k}^2}\,\tanh\left(\frac{3a_{0k}}{\rho}\sqrt{1 - 3a_{0k}^2}\,\xi\right), \quad k = 1,2,3. \tag{35}$$

These are, obviously, kink-type solitons, shown in Fig. 3. We can see that the functions $\psi_k(\xi)$ depend on $\rho$ and $\sigma$ as the parameters $a_{0k}$ depend on $\sigma$. The solitonic width, i.e. its slope, depends on both $\rho$ and $\sigma$, while the jumps of the functions $\psi_k(\xi)$ from $-\infty$ to $+\infty$ depend on $\sigma$ only. Obviously, the solitonic width is proportional to viscosity $\rho$.

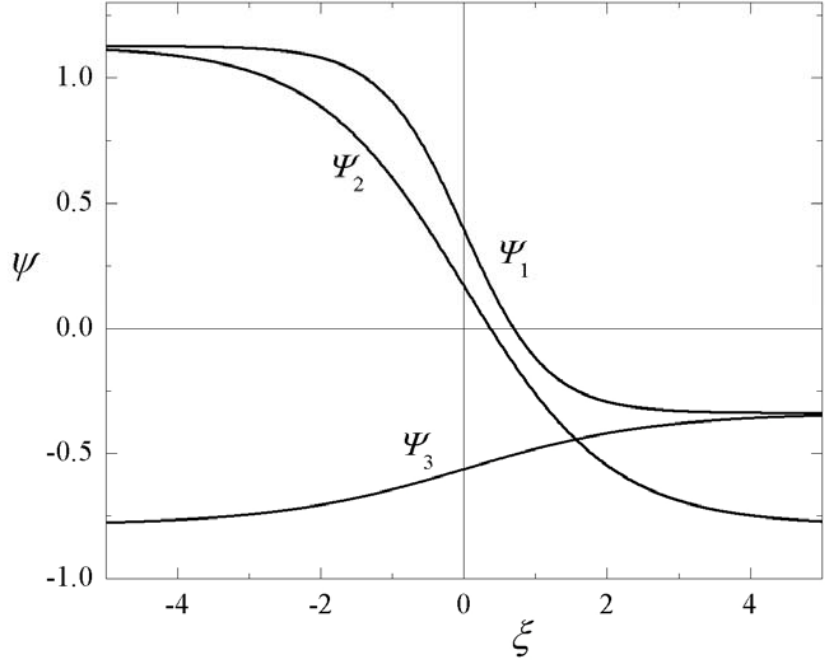

**Fig. 3.** The functions $\psi_k(\xi)$ for $\rho = 1$ and $\sigma = 0.3$.

## 4. Some solutions of Eq. (4) that may not have biophysical meaning

We first investigate the case $\sigma > \sigma_0$. The polynomial given by Eq. (25) has only one real root, which is [27]

$$a_0 = -\frac{1}{\sqrt{3}\sin(2\varphi)}, \qquad \tan\varphi = \sqrt[3]{\tan\left[0.5\arcsin\left(\frac{\sigma_0}{\sigma}\right)\right]} \qquad (36)$$

and the solution of Eq. (4) is

$$\psi_t(\xi) = a_0 + \sqrt{3a_0^2 - 1}\,\tan\left(\frac{3a_0}{\rho}\sqrt{3a_0^2 - 1}\,\xi\right). \qquad (37)$$

Of course, the problem with this solution is the fact that tangent diverges. However, this does not necessarily mean that the solution given by Eq. (37) does not have any physical meaning at all. It is well known that MT is an extremely unstable structure and we speculated recently that the functions including tangent terms might describe its blow up [28].

Notice that $1-3a_0^2 < 0$ if $\sigma > \sigma_0$. This means that both $a_1$ in Eq. (26) and $\beta$ in Eq. (27) are imaginary and, to derive Eq. (37), a formula $\tanh(ix) = i \tan x$ was used.

Our next task is to solve the system of Eqs. (18)-(24) in general, i.e. for $a_1 \neq 0$ and $b_1 \neq 0$. One can easily show that the system reduces to five equations. The solution of this system is given by

$$4a_1^2 = 1 - 3a_0^2, \tag{38}$$

$$a_1 = b_1 \tag{39}$$

and by Eqs. (25), (27) and (28). The final result is

$$\psi_{kn}(\xi) = a_{0k} - \sqrt{1 - 3a_{0k}^2} \coth\left(\frac{3a_{0k}}{\rho}\sqrt{1 - 3a_{0k}^2}\,\xi\right), \quad k = 1,2,3. \tag{40}$$

A next step is to study if the remaining two JEFs can be used for the solution of Eq. (4). We can easily see that all the terms except $\psi'$ are powers of either cn or dn functions. However, the first derivative $\psi'$ cannot be expressed like this for any value of the modulus $m$. This, certainly, means that sn is the only choice for the solution of Eq. (4).

## 5. General case

Equations like Eq. (4) appears rather often. Let us mention just one more example. The model where only longitudinal displacements of dimers are taken into consideration is described in Section 2. If one radial degree of freedom per dimer is assumed we can use the potential energy $-\vec{p}\cdot\vec{E} = -pE\cos\varphi$ instead of the one given by Eq. (2), where $\vec{p}$ and $\vec{E}$ are a moment of the electric dipole and electric field strength, respectively. Hence, $\sin\varphi$ appears in Eq. (3) and a series expansion brings about terms $\varphi$ and $\varphi^3$. Finally, instead of Eq. (4) we obtain [29]

$$\alpha\psi'' - \rho\psi' + \psi - \psi^3 = 0. \tag{41}$$

It is obvious that Eq. (41) is simpler than Eq. (4) and, of course, can be solved using the same method as above. Therefore, it is extremely important to know what kind of ODEs can be solved using JEFs, and this is a topic of this section.

Let us assume that the equation to be solved is

$$\alpha_1\psi^{(n)} + \alpha_2\psi^{(n-1)} + .... + \beta_1\psi^N + \beta_2\psi^{N-1} + ... + C = 0, \tag{42}$$

representing an extension of Eq. (4), where $\psi^{(n)}$ is the nth derivative with respect to $\xi$, and $\alpha_i$, $\beta_i$ and $C$ are real constants. The highest orders of the function $\Phi$ in the expressions for $\psi^{(n)}$ and $\psi^N$ are $\Phi^{M+n}$ and $\Phi^{NM}$, respectively. If the highest derivative is the second one, like above, we come up with $M+2=NM$. This means that the methods explained in the previous two sections can be used only for the two cases. These are $N=3$ ($M=1$) and $N=2$ ($M=2$). If the highest derivative is $n$ we easily obtain

$$M = \frac{n}{N-1}, \qquad (43)$$

representing a combination of the integers. For example, if the highest power is 5 then the methods can be used if $n$ is a multiple of $N-1=4$. In other words, the values $n=4,8,12,...$ require the series expansions with $M=1,2,3,...$, respectively. Of course, the modulus $m=1$ is assumed.

## 5. Conclusion

In this paper, we investigated the nonlinear dynamics of microtubulin system by invoking an analytical method based on one of the JEFs. The method is very simple and powerful but, like other methods, cannot be used to solving any ODE. We explained how we can know if the method can be used or not.

For the example studied in this paper the solutions are given by Eqs. (30)-(35) and by Fig. 3. To understand their meanings better we should keep in mind that the potential given by Eq. (2) is nonsymmetrical, having two minima. The asymptotic values of the functions $\psi_k(\xi)$ are

$$\left.\begin{array}{l} \psi_1(-\infty)=\psi_{\min 1}, \quad \psi_1(+\infty)=\psi_{\max} \\ \psi_2(-\infty)=\psi_{\min 1}, \quad \psi_2(+\infty)=\psi_{\min 2} \\ \psi_3(-\infty)=\psi_{\min 2}, \quad \psi_3(+\infty)=\psi_{\max} \end{array}\right\}. \qquad (44)$$

Hence, these functions represent transitions from the minima to the maximum and from one to the other minimum. From the physical point of view we believe that the best solution is $\psi_1(\xi)$. Namely, we expect the system to be in a deeper minimum, denoted as min1, which is a ground state. When a dimer obtains a portion of energy it may move to the second minimum or to the maximum. In the first case one more portion of energy would be required for the system to return to the initial ground state. On the other hand, if the system is around its maximum it can spontaneously return to the deeper minimum.

**Acknowledgements**

This research was supported by funds from Serbian Ministry of Education and Sciences, grant III45010

L. Kavitha gratefully acknowledges the financial support from UGC, NBHM, India in the form of major research projects, BRNS, India in the form of Young Scientist Research Award and ICTP, Italy in the form of Junior Associateship.

A. Muniyappan gratefully acknowledges UGC for Rajiv Gandhi National Fellowship.

S. Zdravković thanks to Dejan Milutinović who plotted figure 1.

**References**

[1] M. Cifra, J. Pokorný, D. Havelka, O. Kučera, Electric field generated by axial longitudinal vibration modes of microtubule, BioSystems 100 (2010) 122-131.

[2] P. Dustin, Microtubules, Springer, Berlin, 1984.

[3] J.A. Tuszyńsky, S. Hameroff, M.V. Satarić, B. Trpisová, M.L.A. Nip, Ferroelectric Behavior in Microtubule Dipole Lattices: Implications for Information Processing, Signaling and Assembly/Disassembly, J. Theor. Biol. 174 (1995) 371-380.

[4] M.V. Satarić, J.A. Tuszyńsky, Nonlinear Dynamics of Microtubules: Biophysical Implications, J. Biol. Phys. 31 (2005) 487-500.

[5] J.E. Schoutens, Dipole–Dipole Interactions in Microtubules, J. Biol. Phys. 31 (2005) 35-55.

[6] D. Havelka, M. Cifra, O. Kučera, J. Pokorný, J. Vrba, High-frequency electric field and radiation characteristics of cellular microtubule network, J. Theor. Biol. 286 (2011) 31-40.

[7] S. Zdravković, M.V. Satarić, S. Zeković, Nonlinear dynamics of microtubules – A new model, Submitted to Commun. Nonlinear Sci.

[8] P. Drabik, S. Gusarov, A. Kovalenko, Microtubule Stability Studied by Three-Dimensional Molecular Theory of Solvation, Biophys. J. 92 (2007) 394-403.

[9] E. Nogales, M. Whittaker, R.A. Milligan, K.H. Downing, High-Resolution Model of the Microtubule, Cell 96 (1999) 79-88.

[10] M.V. Satarić, J.A. Tuszyński, R.B. Žakula, Kinklike excitations as an energy-transfer mechanism in microtubules, Phys. Rev. E 48 (1993) 589-597.

[11] M.A. Collins, A. Blumen, J.F. Currie, J. Ross, Dynamics of domain walls in ferrodistortive materials. I. Theory, Phys. Rev. B 19 (1979) 3630-3644.

[12] A. Gordon, Nonlinear mechanism for proton transfer in hydrogen-bonded solids, Physica 146 B (1987) 373-378.

[13] A. Gordon, Kink dynamics in hydrogen-bounded solids, Physica B 151 (1988) 453 .

[14] A. Gordon, Conductivity by ionic defects in hydrogen-bonded chains, Solid State Commun. 69 (1989) 1113-1115.

[15] M.V. Satarić, Dj. Koruga, Z. Ivić, R. Žakula, The Detachment of Dimers in the Tube of Microtubulin as a Result of Solitonic Mechanism, J. Mol. Electron. 6 (1990) 63-69.

[16] S.A. Elwakil, S.K. El-labany, M.A. Zahran, R. Sabry, Modified extended tanh-function method for solving nonlinear partial differential equations, Phys. Lett. A 299 (2002) 179-188.

[17] A.H.A. Ali, The modified extended tanh-function method for solving coupled MKdV and coupled Hirota–Satsuma coupled KdV equations, Phys. Lett. A 363 (2007) 420-425.

[18] S.A. El-Wakil, M.A. Abdou, New exact traveling wave solutions using modified extended tanh-function method, Chaos, Solitons and Fractals 31 (2007) 840-852.

[19] L. Kavitha, A. Prabhu, D. Gopi, New exact shape changing solitary solutions of a generalized Hirota equation with nonlinear inhomogeneities, Chaos, Solitons and Fractals 42 (2009) 2322-2329.

[20] L. Kavitha, B. Srividya, D. Gopi, Effect of nonlinear inhomogeneity on the creation and annihilation of magnetic soliton, J. Magn. Magn. Mater. 322 (1010) 1793-1810.

[21] L. Kavitha, N. Akila, A. Prabhu, O. Kuzmanovska-Barandovska, D. Gopi, Exact solitary solutions of an inhomogeneous modified nonlinear Schrödinger equation with competing nonlinearities, Math. Comput. Modelling. 53 (2011) 1095-1110.

[22] M. Remoissenet, Waves Called Solitons, Springer-Verlag, Berlin, Heidelberg, 1989.

[23] N.I. Akhiezer, Elements of the Theory of Elliptic Functions, Translations of mathematical monographs 79, American Mathematical Society, Providence, 1990.

[24] M. Lakshmanan, S. Rajasekar, Nonlinear Dynamics, Springer-Verlag, Berlin, Heidelberg, 2003.

[25] C. Dai, J. Zhang, Jacobian elliptic function method for nonlinear differential-difference equations, Chaos, Solitons and Fractals 27 (2006) 1042-1047.

[26] A. Scott, Nonlinear Science Emergence and Dynamics of Coherent Structures, Fizmatlit, Moscow, 2007. (in Russian).

[27] V.I. Smirnov, Kurs vishey matematiki, tom 1, Nauka, Moscow, 1965. (In Russian).

[28] S. Zdravković, L. Kavitha, M.V. Satarić, S. Zeković, Jovana Petrović, Modified extended tanh-function method and nonlinear dynamics of microtubules, Submitted to Chaos, Solitons and Fractals.

[29] S. Zdravković, M.V. Satarić, A. Maluckov, work in preparation.